\documentstyle[sprocl]{article}

\def\half{{{1\over2}}}

\def\be{\begin{equation}}
\def\bea{\begin{eqnarray}}
\def\eea{\end{eqnarray}}
\def\ee{\end{equation}}
\def\simleq{\; \raise0.3ex\hbox{$<$\kern-0.75em
      \raise-1.1ex\hbox{$\sim$}}\; }
\def\simgeq{\; \raise0.3ex\hbox{$>$\kern-0.75em
      \raise-1.1ex\hbox{$\sim$}}\; }

\def\apj#1#2#3{{\it Astrophys.\ J.\ }{{\bf #1} {#2} {(#3)}}}
\def\app#1#2#3{{\it Astropart.\ Phys.\ }{{\bf #1} {#2} {(#3)}}}
\def\np#1#2#3{{\it  Nucl.\ Phys.\ }{{\bf #1} {#2} {(#3)}}}
\def\pr#1#2#3{{\it Phys.\ Rev.\ }{{\bf #1} {#2} {(#3)}}}
\def\pl#1#2#3{{\it  Phys.\ Lett.\ }{{\bf #1} {#2} {(#3)}}}
\def\prl#1#2#3{{\it Phys.\ Rev.\ Lett.\ }{{\bf #1} {#2} {(#3)}}}
\def\nat#1#2#3{{\it Nature }{{\bf #1} {#2} {(#3)}}}

\begin{document}

\title{Neutrino oscillations in magnetized media \\
and implications for the pulsar velocity puzzle}

\author{D. Grasso}

\address{Departament de F\'isica Teorica, 
        Universitat de Val\`encia \\ 
        E-46100  Burjassot, Val\`encia, SPAIN}

\maketitle\abstracts{
After a brief presentation of the general techniques used to determine 
neutrino potentials in a magnetized medium I will discuss  
MSW resonant oscillations of active and sterile neutrinos in such environment. 
Using my results I will reconsider the viability of a solution of the 
pulsar velocity puzzle based on such a kind of neutrino oscillations.}


\section{Introduction}

The effect of media on the propagation of neutrinos is
nowadays a well established research subject \cite{Raffelt}. 
Among the several relevant implications of this subject for particle physics and
astrophysics, perhaps the  most remarkable application was the discovery of the 
possibility to have neutrino resonant oscillations in the matter (MSW) and 
the study of the possible role that this effect may play to give a solution 
to the solar-neutrino problem \cite{MSW}.

Recently it has been proposed that MSW oscillation may help to solve 
another astrophysical puzzle, namely that of the high velocity 
of pulsars \cite{KusSeg96}. 
Observations \cite{veloc} show that these velocities range from zero 
up to 900 km/s with a mean value of $450 \pm 50$ km/s . 
Among several models, the possibility that pulsar motion is a consequence of 
an asymmetric neutrino emission during the supernova (SN) explosion is 
particularly
attractive. In fact, neutrinos carry more than $99 \%$ of the SN
gravitational binding energy so that even a $1 \%$ asymmetry in the neutrino
emission could generate the observed pulsar velocities.
The strong magnetic field present during a SN explosion may induce such an
asymmetry thanks to its interplay with the parity violating weak interaction
in which neutrinos are involved. That this effect may actually take place
in collapsed stars has been observed by several authors 
(see e.g. \cite{Chugai}).     

In this contribution I will discuss another realization of this  
idea based on neutrino resonant oscillations.

\section{Neutrino potentials in a magnetized medium}

The main tool to study neutrino propagation in a medium, like in the vacuum,
is the Dirac equation. In the Fourier space this reads
\be
\label{Dirac}
\left( \partial_\mu \gamma^\mu - m_\nu - \Sigma(T,B,\mu_i) \right) \psi_\nu(k) 
= 0~.
\ee
Here $\Sigma(T,B,\mu_i)$, where $T$ is the heath-bath temperature, $\mu_i$ the
chemical potential of the $i$-th particle species and $B$ the magnetic field
strength, is the medium induced self-energy. The vacuum contribution to 
$\Sigma$ is already hidden in the neutrino physical vacuum mass $m_\nu$. 
Two kind of Feynman diagrams contribute to $\Sigma(T,B,\mu_i)$. These are
the bubble and the tadpole diagrams discussed in ref.\cite{EGR}.
The fermions running in the
internal lines of these diagrams have to be intended as real (not virtual)
particles belonging to the heat-bath. At the temperatures and densities 
generally present in SNs we can safely neglect any thermal population of
gauge-bosons, so that $W$ and $Z$ in ours diagrams are the usual virtual
ones.  Thermal populations of muons and tauons are also negligible, so that
only electrons, positrons , proton and neutrons give a relevant contribution
to $\Sigma(T,B,\mu_i)$.
If we neglect (see below) nucleon polarizations, the ambient magnetic
field can affect $\Sigma$ only through its effect on the electron propagator.
A very powerful tool to study this effect is given by the finite-temperature 
electron propagator in the presence of an external magnetic field.
This propagator has been derived in refs. \cite{Mak} and
applied for the first time to neutrino physics in \cite{EGR}. 
As the complete expression of this propagator is quite long I do not 
report it here. The interested reader can find it in  refs. \cite{EGR,Mak}.
With such a  tool a complete one-loop expression for $\Sigma(T,B,\mu_i)$ 
can be determined \cite{EGR}. 
Inserting such a result into the neutrino dispersion relation
\be
\det\left( \partial_\mu \gamma^\mu - m_\nu - 
\Sigma(T,B,\mu_i) \right) = 0
\ee
we can then derive the matter induced potential of neutrinos propagating
through an electrically neutral plasma 
\bea
\label{ve}
V(\nu_e) &=& \sqrt{2} G_{\rm F} \left[ - \frac{N_n}{2} +  N_e + 2 N^L_{\nu_e} 
+ \sum_{i=\mu,\tau}\!N^L_{\nu_i}  -  \half N^0_e\, \cos\phi \right]  \\
\label{vt}
V(\nu_{\mu,\tau}) &=&  \sqrt{2} G_{\rm F} \left[ - \frac{N_n}{2} +
\sum_{i=e,\mu,\tau}\!N^L_{\nu_i} +
N^L_{\nu_{\mu,\tau}} +  \half N^0_e\, \cos\phi \right]~.
\eea
where $\phi$ is the angle between neutrino wave-vector and the 
magnetic field vector.
In my notation $N_i$ represents the charge density of the $i$-th particle 
species.  $N^0_e$ stands for the electron charge density in the lowest Landau 
level (LLL). 
Both  quantities $N_e$ and $ N^0_e$ are functions of $T, \mu_i$ and ${\vec B}$
and increase almost linearly with $B$ when $eB \gg \mu_e^2, T^2$.
Since, due to the double degeneracy of the $n \ge 1$ Landau level, only
the LLL contributes to the spin-polarization of the electron-positron gas,
it is possible \cite{NSSV} to rewrite the angular dependent terms in 
(\ref{ve},\ref{vt}) in terms of a polarization parameter defined by 
$\lambda \equiv  N^0_e/ N_e$. 

The reader should keep in mind that nucleon polarizations may also
induce  angular dependent terms in the neutrino potentials.    
Although nucleon anomalous magnetic moments are much smaller than the 
electrons', it is however possible that a strong degeneration of the electron
gas suppress electron polarization. If, at same time, the nucleons are
non-degenerate it may then happen that nucleon polarization is not negligible
\cite{NSSV,ALS}. A similar situation may actually be realized in the 
central regions of hot NSs. However, the effect of nucleons 
polarization on MSW oscillations is generally subdominant.

To conclude this section we summarise that the  effect of 
the magnetic field on the neutrino propagation is  two-folds. 
Strong magnetic fields may modify the bulk properties
of the electron-positron gas inducing an enlargement of the potential 
to which  the $\nu_e$ is submitted. 
Besides that, the magnetic field induces a polarization of particles
having a non-vanishing magnetic moment. The latter effect
gives rise to an angular dependence in the neutrino potential.    

I finally observe that expressions  $V(\nu_e)$ and $V(\nu_{\mu,\tau})$ disagree 
with those reported in \cite{EspCap} and \cite{KusSeg97} whereas they are 
consistent with those given in \cite{NSSV,ALS}.

\section{Resonant oscillations}

I will now discuss some implications of the result reported in the previous 
section for MSW-type neutrino resonant oscillations.
The general form of resonance condition for such a kind of oscillations is
\be
\label{genres}
V(\nu_i) + \frac{m_{\nu_i}}{2E} = V(\nu_{j}) + 
\frac{m_{\nu_j}}{2E}~. 
\ee
In the case of $\nu_e - \nu_{\mu,\tau}$ oscillations,
substituting (\ref{ve},\ref{vt}) in (\ref{genres}) we find
(see also \cite{NSSV} for an independent derivation of the same result)
\be
\label{reson}
\frac{\Delta m^2}{2E} \cos2\theta =  \sqrt{2} G_{\rm F} N_e \left( 1 -
\lambda \cos\phi \right)
\ee
where $\Delta m^2 \equiv m_{\nu_{\mu,\tau}}^2 -  m_{\nu_e}^2$
and $\theta$ is the vacuum mixing angle. 
Since in a SN the charge asymmetry of neutrinos is usually
negligible, I have ignored here the heath-bath neutrino 
contribution to the potential.

Resonant MSW oscillations between active and sterile neutrinos can also
be studied. As $V(\nu_S) = 0$ we find the resonance conditions to be
\bea
\label{e-s}
\frac{\Delta m^2}{2E}\cos2\theta &=& \sqrt{2} G_{\rm F} 
\left[ N_e \left( 1 + \half \lambda \cos\phi \right) 
- \half N_n \right] \\
\label{tau-s}
\frac{\Delta m^2}{2E}\cos2\theta &=& \sqrt{2} G_{\rm F} 
\left[ N_e \left( \half \lambda \cos\phi \right) 
- \half N_n \right]
\eea
respectively for $\nu_e \leftrightarrow \nu_S$ and
 $\nu_{\mu,\tau} \leftrightarrow \nu_S$ oscillations.
Here $\Delta m^2 \equiv m_{\nu_S}^2 - m_{\nu_{e,\mu,\tau}}^2$. 
Whenever  $N_e \ll \half N_n$ (typically below the $\tau$-neutrinosphere),
we see from (\ref{tau-s})
that  $\nu_{e} \leftrightarrow \nu_S$ resonant oscillations are
only possible if $\Delta m^2 < 0$. On the contrary, if $\Delta m^2 > 0$ only
${\bar \nu}_{e} \leftrightarrow {\bar \nu}_S$ resonant oscillations 
are possible. Note that in such a case the asymmetry in the resonant surface
would be rotated by $\pi$ with respect to the $\nu_e \leftrightarrow \nu_S$ 
case. 
Concerning  $\nu_{\mu,\tau} \leftrightarrow \nu_S$ resonant oscillations,
they are only possible for a negative value of  $\Delta m^2$ or, in the 
opposite case, only between the respective anti-particles.

\section{Consequences for pulsar velocities}

The possibility that the angular dependence of the neutrino resonance
surface in SNs could be the origin of the observed velocities of pulsars
has been first proposed by Kusenko ansd Segre \cite{KusSeg96}.
To this purpose, Kusenko ansd Segre started considering  $\nu_e-\nu_\tau$ 
oscillations.
Their idea bears on the assumption that the resonance sphere lies 
between the $\tau$ and $e$-neutrinospheres. In fact, if this is the case
the distortion of the resonance sphere would induce a
temperature anisotropy of the escaping $\tau$-neutrinos produced by the
oscillations, hence a recoil kick of the proto-neutron star.
In order to account for the observed pulsar velocities a $1 \%$ asymmetry 
in the escaping neutrino total momentum is required. 

Kusenko and Segre computed the asymmetry in the $\tau$-neutrino flux
in a weak field limit and assuming the electron to neutron number density 
ratio to be uniform close to the neutrino-spheres.
However, such conditions are generally not fulfilled in a SN environment
and a more general treatment is called for.
Especially, electron relative abundance decreases steeply close to the
neutrinospheres as a consequence of the strong deleptonization taking place
in that region during the Kelvin-Helmholtz cooling phase.
Furthermore, as a consequence of such an effect $eB$ may become larger or 
comparable to the electron Fermi momentum squared in this region, so that 
LL quantization cannot be ignored.
Following \cite{Qian} and by using (\ref{reson})
we determine the asymmetry in the escaping $\tau$-neutrino
momentum to be
\be
\label{asy}
\frac{\Delta k}{k} \approx \frac{1}{6} \frac
{ \int_0^\pi F_{\nu} \cos\phi \sin\phi d\phi}
{ \int_0^\pi F_{\nu} \sin\phi d\phi} \simeq
\frac{2}{9} \frac{h_{N_e}}{h_T} \lambda
\ee
where $F_\nu$ is the flux of the neutrino produced by the resonant oscillations and, 
$h_{N_e} \equiv |d\ln N_e/dr|^{-1}$ and $h_T \equiv |d\ln T/dr|^{-1}$
are, respectively, the variation scale heights of $T$ and $N_e$ at the
resonance mean radius. 
The result reported in (\ref{asy}) is more general of that reported in \cite{Qian}
as it applies also to those case for which  $eB \simleq T^2,\!\mu^2$.
Between the two neutrinospheres the value of the ratio 
$\frac{h_{N_e}}{h_T}$ is typically 
$\approx 1$, the exact value depending on the adopted SN model. 
We then see from (\ref{asy}) that at least a $\sim 10 \%$ electron 
polarization is required to achieve the desired anisotropy.  
This requirement translates (see e.g. ref.\cite{NSSV})
into a minimal value of the dipolar component of the
magnetic field strength which lies in the range $10^{15}\div 10^{16}$ Gauss.
This value is larger of at least one order of-magnitude than that claimed in 
\cite{KusSeg96}.
In order for the resonance surface to lie between the two neutrino-spheres
the $\nu_\tau$ mass has to be of the order of $100$ eV.
This would be at odd with cosmological constraints unless $\nu_\tau$ is
unstable.

Active-sterile resonant neutrino oscillation may also play a role in 
accelerating pulsars. As the radius of the sterile-neutrinosphere is zero,
in order to produce a recoil kick in this case we just need
 the resonance surface to be placed within the neutrinosphere of the active 
neutrino participating in the  oscillations. The possible role that
${\bar \nu}_{\mu,\tau} \leftrightarrow {\bar \nu_S}$  resonant oscillations 
may have in this context was considered,   in \cite{KusSeg97}.
As follows from (\ref{tau-s}), a value of the neutrino squared mass difference
 $\Delta m^2 \approx 1^2\  {\rm keV}^2$ is, in this case, required. 
${\nu}_{e} \leftrightarrow {\nu_S}$ oscillations were 
instead ignored in \cite{KusSeg97} as the $V(e)$  angle dependence
was  erroneously disregarded in that work.
A solution of the pulsar velocity puzzle based on this kind of oscillations is,
 however, still viable and appealing as it does not enter in conflict with 
cosmological bounds.  In fact, using (\ref{e-s}) we can write the equation
determining the mean resonance position as
\be
\label{rese}
\frac{\Delta m^2}{2 E} = \frac{G_F \rho}{\sqrt{2}m_N} \left(3Y_e - 1 \right)\ ,
\ee 
where, using electric charge neutrality,  $Y_e \equiv N_e/(N_p + N_n) = 
1 - Y_n$ and $m_N$ is the nucleon mass.
We see from (\ref{rese}) that resonant oscillation may take place even
for very small values of $\Delta m^2$ (eventually even compatible with a MSW 
inspired solution of the solar neutrino problem) provided that 
$Y_e \simeq Y_n/1 \approx 1/3$. Indeed, this condition is expected to be 
fulfilled for $\rho \approx 10^{12}$, that is in proximity of the 
$e$-neutrinosphere \cite{Voloshin,ALS}.
The required magnetic field strength is, also in this case, in the range  
$10^{15}\div 10^{16}$ Gauss.

We conclude that MSW oscillations may provide
a very elegant solution to the problem of the origin of pulsar velocities.
Although such a mechanism generally requires strong dipolar   
magnetic fields to be present in the inner core of SNs
(even larger of what claimed in previous works), still,
present observations does not exclude this intriguing possibility.

\section*{Acknowledgments}
This work was supported by DGICYT under grant number PB95-1077 and by
the TMR network grant ERBFMRXCT960090.


\section*{References}
\begin{thebibliography}{99}

\bibitem{Raffelt} G. Raffelt, Stars as a Laboratory for Fundamental Physics, 
The University of Chicago Press, 1996.

\bibitem{MSW} L. Wolfenstein, \pr{D17}{2369}{1978};
S.P. Mikheyev and A.Y. Smirnov, {\it Nuovo Cimento} {\bf 9C}, 17 (1986).

\bibitem{KusSeg96} A. Kusenko and G. Segr\`e, \prl{77}{4872}{1996}.

\bibitem{veloc} A.G. Lyne and D.R.~Lorimer, \nat{369}{127}{1994}.

\bibitem{Chugai} N.N. Chugai, {\it Pis'ma Astron. Zh.} {\bf 10}, 87 (1984);
A. Vilenkin, \apj{451}{700}{1995}.

\bibitem{EGR} P.~Elmfors, D.Grasso and G. Raffelt, \np{B479}{3}{1996}.

\bibitem{Mak} K.W. Mak, \pr{D49}{6939}{1994}; 
P. Elmfors, D. Persson and B.-S. Skagerstam, \app{2}{299}{1994}.

\bibitem{NSSV} H. Nunokawa, V.B. Semikoz, A.Y. Smirnov and J.W.F. Valle,
\np {B501}{17}{1997}.

\bibitem{ALS} E.Kh. Akhmedov, A. Lanza and D.W.Sciama, \pr{D56}{6117}{1997}.

\bibitem{EspCap} S. Esposito and G. Capone, {Z. Phys.} {C70}, 55 (1996).

\bibitem{KusSeg97} A. Kusenko and G. Segr\`e, \pl{B396}{197}{1997}.

\bibitem{Qian} Y.Z. Qian, \prl{79}{2750}{1997}.

\bibitem{Voloshin} M.B. Voloshin, \pl{B209}{360}{1988}.


\end{thebibliography}
\end{document}